\documentstyle[12pt]{article}

\begin{document}

\setcounter{page}{1}

\pagestyle{plain} \vspace{1cm}
\begin{center}
\Large{\bf Reissner-Nordstr\"{o}m Black Hole Thermodynamics in Noncommutative Spaces}\\
\small \vspace{1cm}
{\bf Kourosh Nozari}\quad \quad and \quad \quad {\bf Behnaz Fazlpour }\\
\vspace{0.5cm} {\it Department of Physics,
Faculty of Basic Science,\\
University of Mazandaran,\\
P. O. Box 47416-1467,
Babolsar, IRAN\\
e-mail: knozari@umz.ac.ir}
\end{center}
\vspace{1.5cm}
\begin{abstract}
This paper considers the effects of space noncommutativity on the
thermodynamics of a Reissner-Nordstr\"{o}m black hole. In the first
step, we extend the ordinary formalism of Bekenstein-Hawking to the
case of charged black holes in commutative space. In the second step
we investigate the effect of space noncommutativity on the
thermodynamics of charged black holes. Finally we compare
thermodynamics of charged black holes in commutative space with
thermodynamics of Schwarzschild black hole in noncommutative space.
In this comparison we explore some conceptual
relation between charge and space noncommutativity.
\end{abstract}
\newpage
\section{Introduction}
Generally, black holes can be characterized by three (and only
three) quantities: mass (M), electric charge (Q) and angular
momentum ($\vec{J}$) (for a review see [1,2,3]). A charged black
hole is a black hole that possesses electric charge. Since the
electromagnetic repulsion in compressing an electrically charged
mass is dramatically greater than the gravitational attraction (by
about 40 orders of magnitude), it is not expected that black holes
with a significant electric charge will be formed in nature. When
the black hole is electrically charged, the Schwarzschild solution
is no longer valid. In this case the Reissner-Nordstr\"{o}m geometry
describes the geometry of empty space surrounding a charged black
hole.

If the charge of black hole is less than its mass (measured in
geometric units G = c = 1), then the geometry contains two horizons,
an outer horizon and an inner horizon. Between the two horizons
space is much like a waterfall, falling faster than the speed of
light, carrying everything with itself. It should be stressed that
fundamental charged particles such as electrons and quarks are not
black holes: their charge is much greater than their respective
masses, and they do not contain horizons.

The issue of black hole thermodynamics and its quantum gravitational
correction has been studied extensively[4,5,6,7]. Since this problem
is a key attribute of quantum gravity proposal, investigation of its
various aspects will shed light on the perspectives of ultimate
quantum gravity scenario. It has been revealed that quantum
corrections to the Bekenstein-Hawking formalism of black hole
thermodynamics can be performed in several alternative approaches
such as noncommutative geometry[6,8], the generalized uncertainty
principle(GUP)[7,9,10] and modified dispersion relations[4,5,11,12].
The goal of the present paper is to proceed one more step in this
direction. we consider the case of charged black holes. We firstly
give an overview to the original formalism of Bekenstein-Hawking for
charged black holes in commutative space. Then we consider the
effects of space noncommutativity on the thermodynamical quantities
of charged black holes. We compare Reissner-Nordstr\"{o}m black hole
thermodynamics in commutative space with thermodynamics of
Schwarzschild black hole in noncommutative space. In this manner we
are forced to conclude that space noncommutativity has some thing to
do with charge. In another words, it seems that space
noncommutativity and charge have the same effects on thermodynamics
of a Schwarzschild black hole.

\section{Charged Black Holes}
By the Schwinger effect in the presence of a charged black hole,
there are pair-creation of charged particles[13]. When we consider
the quantum effects, a charged matter fluid will surround the
singularity and then black hole charge will be screened. Therefore
we will have an electric field which modifies the geometry of the black hole.\\
Consider the Reissner-Nordstr\"{o}m geometry, describing a static
electrically charged black hole with the following line-element
\begin{equation}
ds^2= f(r) dt^2 -\frac{dr^2}{f(r)}- r^2 \left( d \vartheta^2 +\sin^2
\vartheta d\varphi^2 \right),
\end{equation}
where
\begin{equation}
 f(r)=1-\frac{2M}{r}+\frac{Q^2}{r^2}.
\end{equation}
This expression has been written in geometric units, where the speed
of light and Newton's gravitational constant are set equal to unity,
c = G = 1. This metric has two possible horizon which can be found
as follows
\begin{equation}
r=M \pm \sqrt{M^2-Q^2}
\end{equation}
These two values are corresponding to the outer and inner horizons.
Therefore, when a black hole becomes charged, the event horizon
shrinks, and another one appears, near the singularity. The more
charged the black hole is, the closer the two horizons are. As more
and more electric charge is thrown into the black hole, the inner
event horizon starts to get larger, while the outer horizon starts
to shrink. The maximum possible charge on the black hole is when the
two horizons come together and merge. If one tried to force in more
charge, both event horizons would disappear, leaving a naked
singularity. Since in the limit of $Q=0$ we should recover the
Schwarzschild radius $r_s=2M$, we consider the plus sign in (3) and
the corresponding radius will be radius of the outer horizon.\\
If we set $M=\frac{r_s}{2}$ in equation (3), we find
\begin{equation}
r=\frac{r_s}{2} \pm \sqrt{\frac{{r_s}^2}{4}-Q^2}=\frac{1}{2}
\Big(r_s \pm \sqrt{{r_s}^2-4 Q^2}\Big)
\end{equation}
In which follows we consider only the outer radius which is
corresponding to the radius of the Schwarzschild black hole. In this
manner, equation (4) can be rewritten as follows
\begin{equation}
r=r_s -\frac{Q^2}{r_s}-\frac{Q^4}{{r_s}^3},
\end{equation}
where we have considered only three first terms of the right hand
side since $Q<M$. When $Q=0$, we get $r=r_s$ which is corresponding
to the event horizon radius of Schwarzschild black hole.\\
After a brief overview of the Reissner-Nordstr\"{o}m black holes, we
apply the original Bekenstein-Hawking formalism to this type of
black holes.

\section{Thermodynamics of a charged Black Hole}
The Hawking temperature of the Schwarzschild black hole is given by
\begin{equation}
T=\frac{M}{2\pi r^2}
\end{equation}
where substitution of $r$ from equation (5) leads to the following
generalized statement
\begin{equation}
T \simeq \frac{M}{2\pi} \Big(r_s
-\frac{Q^2}{r_s}-\frac{Q^4}{{r_s}^3}\Big)^{-2}
\end{equation}
which leads to the following relation( note that $Q<M$)
\begin{equation}
T \simeq \frac{1}{8\pi M} \Big(1+ \frac{Q^2}{2
M^2}+\frac{5}{16}\frac{Q^4}{M^4}\Big).
\end{equation}
Now we calculate entropy of the charged black hole. In the standard
Bekenstein argument, the relation between energy and position
uncertainty of a particle in the vicinity of black hole event
horizon is given by $ E \geq \frac{1}{\delta x}$ [4]. We suppose
$\delta x\sim r$ and therefore, we find the following generalization
\begin{equation}
E \geq \frac{1}{\Bigg(r_s
-\frac{Q^2}{r_s}-\frac{Q^4}{{r_s}^3}\Bigg)}.
\end{equation}
This relation implicitly shows the necessary modification of the
standard dispersion relations. These modified dispersion relations
have been appeared in scenarios such as loop quantum gravity where
Lorentz invariance violation has been encountered[4,14]. Now
consider a quantum particle that starts out in the vicinity of an
event horizon and then ultimately absorbed by black hole. For a
black hole absorbing such a particle the minimal increase in the
horizon area can be expressed as $(\Delta A)_{min}\geq
4(\ln2)E\delta x$[4]. In this situation, the increase of the event
horizon area can be given as follows
\begin{equation}
\Delta  A \geq 4 (\ln 2)\frac{1}{\Bigg(1
-\frac{Q^2}{{r_s}^2}-\frac{Q^4}{{r_s}^4} \Bigg)},
\end{equation}
where $\ln 2$ is the calibration factor. This statement leads us to
the following relation
\begin{equation}
\frac{dS}{d A} \approx \frac {\Delta S_(min)}{\Delta
A_(min)}\simeq\frac {\ln2}{4(\ln2)\frac{1}{\Bigg(1-\frac{Q^2}
{{r_s}^2}-\frac{Q^4}{{r_s}^4}\Bigg)}}.
\end{equation}
Therefore we can write
\begin{equation}
\frac {dS}{d A}\simeq\frac{1}{4}\Bigg[1
-\frac{Q^2}{{r_s}^2}-\frac{Q^4}{{r_s}^4}\Bigg].
\end{equation}
Now we should calculate $d A$. Since $A =4\pi r^2$, we find
\begin{equation}
A=A_s -8 \pi Q^2-\frac{(4 \pi)^2 Q^4}{A_s}+2 \frac{(4 \pi)^3
Q^6}{{A_s}^2},
\end{equation}
and therefore
\begin{equation}
d A =\Bigg[1+\Big(\frac{4\pi }{A_s}\Big)^{2} Q^4
-4\Big(\frac{4\pi}{A_s}\Big)^{3}Q^6\Bigg]d A_s,
\end{equation}
where $A_s = 4\pi r_{s}^{2}$. Integration of (12) leads to the
following result
\begin{equation}
S \simeq \frac{A_s}{4}-\pi Q^2 \ln{\frac{ A_s}{4}}+ \frac{1}{3}(\pi
Q^{2})^{2}\Big(\frac{4}{A_{s}}\Big)+\frac{5}{2} (\pi
Q^2)^{3}\Big(\frac{4}{ A_s}\Big)^2+{\cal O} \Big((\frac{4}{
A_s})^3\Big)...,
\end{equation}
If we calculate corrections of all orders, we will arrive at the
following compact and generalized form for entropy of
Reissner-Nordstr\"{o}m black holes in commutative spaces
\begin{equation}
S= \frac{A_s}{4}-\pi Q^2 \ln{\frac{
A_s}{4}}+\sum_{n=1}^{\infty}c_{n}\Big(\frac{4}{A_s}\Big)^{n}+\cal{C}
\end{equation}
where $\cal{C}$ is a constant and $c_{n}$'s are quantum gravity
model dependent coefficients. A similar expression for entropy can
be obtain in other alternative approaches such as string theory and
loop quantum gravity. In the case of $Q=0$ this equation yields the
standard Bekenstein entropy, $S= \frac{A_s}{4}$.

This is a generalization of Bekenstein-Hawking formalism to the case
of charged black holes in commutative space. In which follows, we
consider the effects of space noncommutativity on the
Bekenstein-Hawking formalism of charged black holes.

\section{The Effect of Space Noncommutativity}
A noncommutative space can be realized by the
coordinate operators satisfying[15,16,17]
\begin{equation}
\left[{\hat x}_i,{\hat x}_j \right] = i \theta_{ij},\;\;\;i,j=1,2,3,
\end{equation}
where $\hat x_{i}$'s are the coordinate operators and $\theta_{ij}$
is a real, antisymmetric and constant tensor, which determines the
fundamental cell discretization of space much in the same way as the
Planck constant $\hbar$ discretizes the phase space. It has the
dimension of $(length)^2$. Canonical commutation relations in
noncommutative spaces read (with $\hbar=1$ )
\begin{equation}
\left[ {\hat x}_i,{\hat x}_j \right]=i \theta_{ij},\;\;\; \left[
{\hat x}_i,{\hat p}_j \right]=i \delta_{ij},\;\;\; \left[ {\hat
p}_i,{\hat p}_j \right]=0,
\end{equation}
There is a new coordinate system with the following definitions
\begin{equation}
\ x_i={\hat x}_i + \frac{1}{2} \theta_{ij} {\hat p}_j,\;\;\;
p_i={\hat p}_i.
\end{equation}
With these new variables, $x_i$'s and $p_i$'s satisfy the
usual(commutative) commutation algebra
\begin{equation}
\ [x_i,x_j] = 0,\;\;\;[x_i,p_j] = i\delta_{ij},\;\;\;[p_i,p_j] = 0.
\end{equation}
In which follows we develop the arguments of the preceding section
to the case where space noncommutativity is present. For a
noncommutative charged black hole, we have
\begin{equation}
f(r)=1 - \frac{2M}{\sqrt{{\hat r}{\hat r}}}+\frac{Q^2}{{\hat r}^2},
\end{equation}
where ${\hat r}$ satisfies conditions (19). The horizon of the
noncommutative metric as usual satisfies the condition ${\hat
g}_{00}=0$, which leads to[18,19]
\begin{equation}
1 - \frac{2M}{\sqrt{{\hat r}{\hat r}}}+\frac{Q^2}{{\hat r}^2}=0 .
\end{equation}
By a coordinate transformation from ${\hat x}_i$ to $x_i$ and then
using relation (19), one can show that horizon of the noncommutative
charged black hole satisfies the following approximate condition
\begin{equation}
1 -
\frac{2M}{\sqrt{(x_i-\frac{\theta_{ij}p_j}{2})(x_i-\frac{\theta_{ik}p_k}{2})}}+
\frac{Q^2}{(x_i-\frac{\theta_{ij}p_j}{2})(x_i-\frac{\theta_{ik}p_k}{2})}=0
.
\end{equation}
This leads us to the following relation
$$1-\frac{2M}{r} \left( 1 + \frac{x_i \theta_{ij} p_j}{2r^2}
 - \frac{\theta_{ij} \theta_{ik} p_j p_k}{8r^2}+\frac{3}{8}
\frac{(x_i \theta_{ij} p_j)^2}{r^4}  \right) +$$
\begin{equation}
\frac{Q^2}{r^2}\left( 1 + \frac{x_i \theta_{ij} p_j}{r^2} -
\frac{\theta_{ij} \theta_{ik} p_j p_k}{4r^2}+\frac{(x_i \theta_{ij}
p_j)^2}{r^4} \right) +{\cal O} (\theta^3) +...=0,
\end{equation}
where $\theta_{ij}=\frac{1}{2} \epsilon_{ijk} \theta_k$. Using the
identity $ \epsilon_{ijr} \epsilon_{iks}= \delta_{jk} \delta_{rs} -
\delta_{js} \delta_{rk}$, one can rewrite (24) as follows

$$1 - \frac{2M}{r} \left[1+\frac{ {\vec L}.{\vec \theta}}{4
r^2} - \frac{\left( p^2 \theta^2 -({\vec p}.{\vec \theta})^2
\right)}{32 r^2} +\frac{ 3({\vec L}.{\vec \theta})^2 }{32 r^4 }
\right]+$$
\begin{equation}
\frac{Q^2}{r^2} \left[1+\frac{ {\vec L}.{\vec \theta}}{2 r^2} -
\frac{\left( p^2 \theta^2 -({\vec p}.{\vec \theta})^2 \right)}{16
r^2} +\frac{ ({\vec L}.{\vec \theta})^2 }{4 r^4 } \right]+ {\cal
O}(\theta^3)+ ...=0,
\end{equation}
where $L_k=\epsilon_{ijk} x_i p_j$,  $p^2={\vec p}.{\vec p}\,$ and
$\,\theta^2={\vec \theta}.{\vec \theta}$ . If we set
$\theta_3=\theta$ and assuming that remaining components of $\theta$
all vanish (which can be done by a rotation or a re-definition of
the coordinates), then ${\vec L}.{\vec \theta}=L_z \theta$ and
${\vec p}.{\vec \theta}=p_z \theta$. Since Reissner-Nordstr\"{o}m
black hole is non-rotating, we set ${\vec L}=0 $ and therefore
$L_z=0$. In this situation equation (25) can be written as
\begin{equation}
r^4 - 2M r^3 +\frac{M \left( p^2 - p_z^2 \right)\theta^2}{16} r+Q^2
r^2-\frac{Q^2 \left( p^2 - p_z^2 \right)\theta^2}{16}  + {\cal
O}(\theta^3)+...=0 .
\end{equation}
From this equation one can see that space noncommutativity has no
effect on Reissner-Nordstr\"{o}m spacetime in first order
approximation. Since $p^2=p_x^2+p_y^2+p_z^2$, one can write $(p^2 -
p_z^2) \theta^2 =(p_x^2+p_y^2) \theta^2$ and therefore (26) can be
rewritten as follows
\begin{equation}
r^4 - 2M r^3 +Q^2 r^2+\frac{M \left( p_x^2 + p_y^2
\right)\theta^2}{16} r-\frac{Q^2 \left( p_x^2 + p_y^2
\right)\theta^2}{16}=0 .
\end{equation}
With the following definitions
\begin{equation}
a \equiv-2M=-r_s,\;\;\;b\equiv Q^2,\;\;\; c \equiv \frac{M \left(
p_x^2 + p_y^2 \right) \theta^2}{16} ,\;\;\;d\equiv-\frac{Q^2 \left(
p_x^2 + p_y^2 \right)\theta^2}{16} ,
\end{equation}
We can write $c=M \alpha=\frac{r_s}{2} \alpha$ and $d=-Q^2
\alpha$, where
\begin{equation}
\alpha=\frac{ \left( p_x^2 + p_y^2 \right)\theta^2}{16}
\end{equation}
Note that $\alpha$ is so defined that contains the effects of space
noncommutativity. Equation (27) has four roots but when $Q=0$ and
spacetime is commutative we should recover $\hat r_s = r_s$.
Therefore, only one root is acceptable which is given by
\begin{equation}
\hat r_s=-a+\frac{b}{a}-\frac{c}{a^2}+\frac{2 b^2}{3
a^3}-\frac{b}{a^3}(A+B)-\frac{4 c b}{3 a^4}+\frac{2c}{a^4}(A+B)
\end{equation}
where
$$A=\frac{2^\frac{1}{3}}{3}(b^2-3ac+12d)\Bigg(2b^3-9abc+27c^2+27a^2d
-72bd$$
\begin{equation}
+\sqrt{-4(b^2-3ac+12d)^3+(2b^3-9abc+27c^2+27a^2d-72bd)^2}\Bigg)^{-\frac{1}{3}}
\end{equation}
and
$$B=\frac{1}{3\times2^\frac{1}{3}}\Bigg(2b^3-9abc+27c^2+27a^2d
-72bd$$
\begin{equation}
+\sqrt{-4(b^2-3ac+12d)^3+(2b^3-9abc+27c^2+27a^2d-72bd)^2}\Bigg)^{\frac{1}{3}}.
\end{equation}
Substitution of the values of $a,b,c$ and $d$ from (28) leads to the
following expression for radius of event horizon in noncommutative
space
\begin{equation}
\hat r_s=r_s-\frac{Q^2}{r_s}-\frac{\alpha}{2 r_s}-\frac{2 Q^4}{3
r_s^3}+\frac{Q^2}{r_s^3}(A+B)-\frac{2 \alpha Q^2}{3 r_s^3}+\frac{
\alpha}{r_s^3}(A+B)
\end{equation}
where now $A$ and $B$ have the following explicit forms
$$A=\frac{2^\frac{1}{3}}{3}(2 Q^4-24 \alpha Q^2+3 \alpha  r_s^2)\Bigg(16 Q^6 -180 \alpha  Q^2 r_s^2+54 \alpha^2
r_s^2+576\alpha Q^4 $$
\begin{equation}
-\sqrt{-4(4 Q^4-48 \alpha Q^2+6 \alpha  r_s^2)^3+(-16 Q^6 +180
\alpha  Q^2 r_s^2-54 \alpha^2 r_s^2-576\alpha
Q^4)^2}\Bigg)^{-\frac{1}{3}}
\end{equation}
and
$$B=\frac{1}{6\times2^\frac{1}{3}}\Bigg(16 Q^6 -180 \alpha  Q^2 r_s^2+54 \alpha^2
r_s^2+576\alpha Q^4$$
\begin{equation}
-\sqrt{-4(4 Q^4-48 \alpha Q^2+6 \alpha  r_s^2)^3+(-16 Q^6 +180
\alpha  Q^2 r_s^2-54 \alpha^2 r_s^2-576\alpha
Q^4)^2}\Bigg)^{\frac{1}{3}}
\end{equation}
If we simplify expressions of $A$ and $B$, we obtain for $A+B$
\begin{equation}
A+B = \sum_{n,m,i}\eta \frac{Q^n \alpha^m
}{r_{s}^i},\;\;\;\;\;\;\;\;\;\;\;\;n,m=0,1,2,... \quad
and\quad\;\;\;i=0,2,4,...
\end{equation}
where $\eta$ is a numerical coefficient. Note that only even powers
of $\frac{1}{r_s}$ appear in this expansion and correspondingly,
only odd powers of $\frac{1}{r_s}$ will appear in (33).\\
To obtain charged black hole thermodynamics in noncommutative space
we proceed in the line of previous section. For simplicity of
calculations, we consider only three first terms of expansions. For
black hole event horizon we have
\begin{equation}
\hat r_s=r_s-\frac{(2 Q^2+ \alpha)}{2 r_s}-\frac{2(
Q^2+\alpha)Q^2}{3 r_s^3}
\end{equation}
Using equations (6) and (37), we obtain the following generalized
statement for temperature of charged black holes in noncommutative
spaces
\begin{equation}
T = \frac{M}{2\pi} \Bigg(r_s-\frac{(2 Q^2+ \alpha)}{2 r_s}-\frac{2(
Q^2+\alpha)Q^2}{3 r_s^3}\Bigg)^{-2}
\end{equation}
which leads to the following relation
\begin{equation}
T = \frac{1}{8\pi M} \Bigg(1+ \frac{(2 Q^2+ \alpha)}{4
M^2}+\frac{\frac{13}{3}( Q^2+\alpha)Q^2+\frac{3}{4} \alpha^2}{16
M^4}\Bigg).
\end{equation}
Now we calculate entropy of charged black hole in noncommutative
space. As previous section with $\delta x=\hat r_s$, we obtain the
following generalization
\begin{equation}
E \geq \frac{1}{\hat r_s},
\end{equation}
which substitution of $\hat r_s$ from (37) leads to the relation
\begin{equation}
E \geq \frac{1}{\Bigg(r_s-\frac{(2 Q^2+ \alpha)}{2 r_s}-\frac{2(
Q^2+\alpha)Q^2}{3 r_s^3}\Bigg)}.
\end{equation}
Again, this relation implicitly shows the modification of standard
dispersion relations in noncommutative spaces. In this manner, the
increase of event horizon area in noncommutative space is given by
\begin{equation}
\Delta  \hat A_s \geq 4 (\ln 2)\frac{1}{\Bigg(1-\frac{(2 Q^2+
\alpha)}{2 r_s^2}-\frac{2( Q^2+\alpha)Q^2}{3 r_s^4} \Bigg)}.
\end{equation}
which leads to the following relation
\begin{equation}
 \frac{dS}{d \hat A_s} \approx \frac {\Delta S (min)}{\Delta
 \hat A_s( min)}\simeq\frac {\ln2}{4(\ln2)\frac{1}{\Bigg(1-\frac{(2 Q^2+
\alpha)}{2 r_s^2}-\frac{2( Q^2+\alpha)Q^2}{3 r_s^4} \Bigg)}}.
\end{equation}
Therefore we can write
\begin{equation}
\frac {dS}{d \hat A_s}\simeq\frac{1}{4}\Bigg[1-\frac{(2 Q^2+
\alpha)}{2 r_s^2}-\frac{2( Q^2+\alpha)Q^2}{3 r_s^4} \Bigg].
\end{equation}
Now we should calculate $d A$. Since $\hat A_s =4\pi \hat r_s^2$, we
find
\begin{equation}
\hat{ A_s}=A_s -4 \pi (2 Q^2+ \alpha)+(4 \pi)^2
\frac{\Big(-\frac{1}{3}(Q^2+ \alpha)Q^2 +\frac{1}{4} \alpha^2
\Big)}{A_s} +\frac{2}{3}(4 \pi)^3 \frac{Q^2\Big(2 Q^4 +3 \alpha Q^2+
\alpha^2 \Big)}{A_s^2}
\end{equation}
and therefore
$$d \hat A_s =\Bigg[1-\Big(\frac{4\pi }{A_s}\Big)^{2}
\Big(-\frac{1}{3}(Q^2+ \alpha)Q^2 +\frac{1}{4} \alpha^2 \Big)$$
\begin{equation}
-\frac{4}{3}\Big(\frac{4\pi}{A_s}\Big)^{3} Q^2\Big(2 Q^4 +3 \alpha
Q^2+ \alpha^2 \Big)\Bigg]d A_s,
\end{equation}
where $A_s = 4\pi r_{s}^{2}$. Integration of (44) leads to the
following relation for entropy of charged black holes in
noncommutative spaces
 $$ S \simeq \frac{A_s}{4}-\pi \frac{(2Q^2+\alpha)}{2} \ln{\frac{ A_s}{4}}+\pi
^{2}\Big(\frac{1}{3}(Q^2+\alpha)Q^2+\frac{1}{4}\alpha^2
\Big)\Big(\frac{4}{ A_s}\Big)$$
\begin{equation}
+\frac{1}{4} \pi ^{3} \Big(6Q^6+9 \alpha Q^4+\frac{5}{2} \alpha^2
Q^2-\frac{1}{4} \alpha^3\Big)\Big(\frac{4}{ A_s}\Big)^2+{\cal O}
\Big((\frac{4}{ A_s})^3\Big)... ,
\end{equation}
Generally, this relation can be written as the following compact
form
\begin{equation}
S\simeq \frac{A_s}{4}-\pi \frac{(2Q^2+\alpha)}{2} \ln{\frac{
A_s}{4}}+\sum_{n=1}^{\infty}c_{n}\Big(\frac{4}{A_s}\Big)^{n}+\cal{C}
\end{equation}
where ${\cal{C}}$ is a constant of integration. Such an event area
dependence of entropy have been obtained in other alternative
approaches such as string theory and loop quantum gravity(see for
example [4,20]. Note that the logarithmic prefactor is a model
dependent quantity[21,22,23,24]. In the case where $Q=0$ and $\alpha
=0$, this expression yields the standard Bekenstein entropy,
\begin{equation}
S\simeq \frac{A_s}{4}.
\end{equation}

\section{The Relation Between Charge and Space Noncommutativity}
As we have shown, thermodynamics of charged black holes in
commutative space can be described with the following equations
\begin{equation}
T = \frac{1}{8\pi M} \Big(1+ \frac{Q^2}{2
M^2}+\frac{5}{16}\frac{Q^4}{M^4}\Big)
\end{equation}
and
\begin{equation}
S= \frac{A_s}{4}-\pi Q^2 \ln{\frac{
A_s}{4}}+\sum_{n=1}^{\infty}c_{n}\Big(\frac{4}{A_s}\Big)^{n}+{\cal{C}}.
\end{equation}
On the other hand, based on a simple analysis much similar to
approach presented in section $4$, one can show that temperature and
entropy of a noncommutative space Schwarzschild black hole are given
as follows[8]
\begin{equation} T \approx \frac{1}{8\pi
M}\Big[1+\frac{\alpha}{4 M^2}-\frac{3 \alpha^2}{8  M^4}\Big]
\end{equation}
and
\begin{equation}
S= \frac{A_s}{4}-\frac{\pi\alpha}{2} \ln{\frac{
A_s}{4}}+\sum_{n=1}^{\infty}c_{n}\Big(\frac{4}{A_s}\Big)^{n}+{\cal{C}}.
\end{equation}
Irrespective of numerical factors which are model dependent,
comparison between equations (50) and (52) suggests that there is a
similarity between the notion of space noncommutativity and the
charge. The same result can be obtained in comparison of (51) and
(53). Therefore, if we accept the universality of black holes
thermodynamics, we can conclude that space noncommutativity has
something to do with charge. In another words, at least in the
spirit of black hole thermodynamics, charge and space
noncommutativity have the same effects.

\section{Summary and Conclusion}
In this paper we have developed formalism of Bekenstein-Hawking to
the case of charged black holes without rotation. A general
statement for entropy of charged black hole has been presented for
this situation. Then we have investigated the effects of space
noncommutativity on the thermodynamics of charged black holes. To
formulate our proposal, we have considered the effect of space
noncommutativity on the radius of event horizons( There is another
point of view which considers the effect of space noncommutativity
on the energy-momentum tensor on the right hand side of Einstein's
equations[6]). In this manner, we have calculated approximate
statements for temperature and entropy of charged black holes in
noncommutative spaces. Our equations show a general mass or event
horizon area dependence much similar to statements which have been
obtained in other alternative approaches[4,5,20,25]. If we accept
that quantum gravitational corrections of Bekenstein-Hawking
formalism have enough generality(as it seems to be the case since
several alternative approaches give the same mass or event horizon
area dependence for temperature and entropy of black holes), then by
comparing equations of charged black holes in commutative spaces
with corresponding equations of Schwarzschild black holes in
noncommutative spaces, we can conclude that charge and space
noncommutativity have close relation. In other words at least their
effects on black hole thermodynamics are the same. We are forced to
conclude that charge can be considered as a source of space
noncommutativity. This issue can be explained as follows: space
noncommutativity comes back to the quantum nature of spacetime at
very short distances(string scale) where fractal nature of spacetime
leads to a minimal observable length scale and therefore the notion
of spacetime fuzziness. On the other hand, when charge is present,
quantum mechanical properties will arise. So, since the origin of
space noncommutativity goes back to quantum properties of spacetime
and these quantum properties can be attributed to charge, one can
relate the notions of space noncommutativity and the charge.

\end{document}